\documentclass{edp-conf}
\usepackage{graphicx,psfig}
\begin{document}

\TitreGlobal{SF2A 2001}

\title{Galaxy Evolution with ALMA}
\author{F. Combes}
\address{Observatoire de Paris, DEMIRM, 61 Av. de l'Observatoire,
F-75 014, Paris, France}
\maketitle
\begin{abstract} 
The submillimeter and millimeter domains are privileged for the exploration of galaxies at high redshift,
because of the negative K-correction: the peak of the dust emission at 60-100 microns is red-shifted in
these domains. Already blind surveys in blank fields with today instrumentation have discovered in the
continuum of the order of one object per square arcmin, with large limitations due to confusion. In the
molecular lines, the K-correction is not so favorable, and it has been until now difficult to detect the
objects, except when they are starbursting monsters, or gravitationally lensed (about a dozen galaxies
have been detected). ALMA will bring more than one order of magnitude improvement, and will not be
affected by confusion. Normal star-forming objects will be detected, and in particular those enshrouded
in dust, so the instrument will be complementary to the optical ones, such as NGST or ELTs.
\end{abstract}
%
\section{Introduction}

\begin{figure}
\centerline{
\psfig{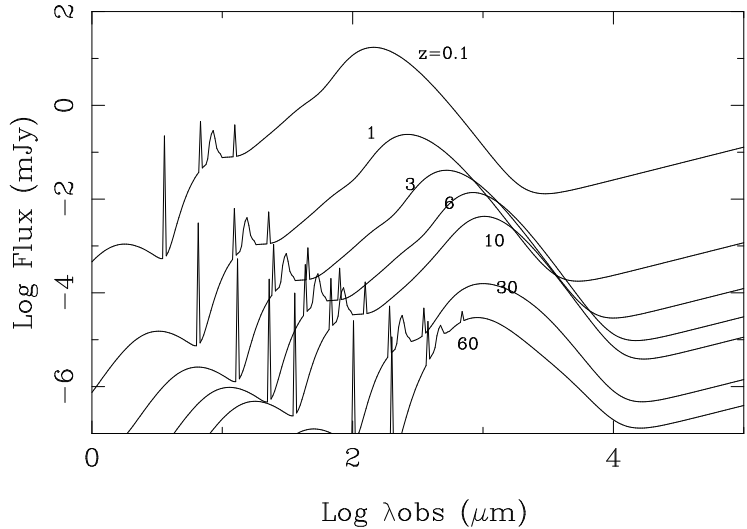}
}
\caption{ SED for a typical starforming galaxy in the radio
and far-infrared, for various  redshifts $z$ = 0.1,
1, 3, 6, 10, 30, 60 ($H_0$ = 75km/s/Mpc, $q_0$ = 0.5).
 The synchrotron spectrum on the right side is a power-law of slope -0.7,
and left the emission from dust is modelled by PAHs, very small
grains and big grains, as in D\'esert et al. (1990) to fit the Milky Way
data. It has been assumed here that the dust properties
are the same as in our Galaxy, and that the power of the starburst
is the same at any redshift.}
\label{sed}
\end{figure}

The formation and evolution of galaxies is one of the main drivers
for the large millimeter array ALMA. This is essentially due to three
arguments:

\begin{itemize}
\item it is now well established that there is a large
increase in the frequency and efficiency of starbursts
at high redshift; there is almost a factor 10 increase from 
$z = 0$ to $z = 1$ in the star formation rate

\item the most active starbursts are the most obscured by dust; this is well
observed at $z = 0$, the young stars are still embedded in their molecular
clouds. At high redshifts, the intense starbursts will almost radiate
all their energy through heated dust. The star formation rate can be highly
under-estimated at optical wavelengths. The far-infrared to optical 
luminosity ratio is observed to vary considerably, between 0.1 and 1000, and
is a good indicator of starbursts

\item the millimeter and sub-millimeter domain are favored
by the negative K-correction: the maximum of dust emission,
that occurs around 100$\mu$m progressively enters in
the millimeter domain (at 0.3mm for $z = 2$). 
The spectral energy distribution (SED) is particularly steep
in this region, since we are in the Rayleigh-Jeans domain,
where the Planck function is B($\nu$) $\propto \nu^2$, and
the optical depth is  $\tau \propto \nu^\beta$, with $\beta \sim$
1.5-2. The emission therefore varies as 4$\sim \nu^4$ (cf figure \ref{sed}).
\end{itemize}

\section{ Present and future surveys at high z}

The present blank field surveys, carried out with 
JCMT-SCUBA or IRAM-MAMBO  at 0.85 -1.2 mm
(Smail et al 1997, Barger et al 1998, Carilli et al 2000)
have been quite successful, and find about 1-2 sources/ arcmin$^2$
above a flux of  1 mJy. The log(N)-log(S) source counts diagram has
been prolonged at low fluxes due to amplification by
gravitational lens effect, through clusters of galaxies.

The main problem encountered by present surveys is the
identification of sources, which has revealed quite difficult.
This is related to the low spatial resolution (of the order of 15''),
which is also source of confusion. The next spatial instruments,
(such as SIRTF, Herschel) will also suffer from confusion, while ALMA
will solve the problem. Another way to identify sources is to obtain
the redshifts through molecular millimetric lines, and this will
be possible with wide-band receivers, such those considered
for the ``redshift machines'' on LMT, GBT, ALMA...

The results of present submm surveys allow to estimate the
star formation rate as a function of redshift, and in particular to 
determine how much this rate is understimated by optical
surveys. After many controversies and debates, the exctinction
correction to apply to visible/UV data is converging towards a factor
about 3 (e.g. Genzel \& Cesarsky 2000), and the star formation rate
is slightly decreasing at high redshift after a maximum at $z =2$.
This estimation is also compatible with the constraints of the
Cosmic Infrared Background (CIB)
(Puget et al 1996, Hauser et al 1998).

A large fraction of the energy produced in starburst is reprocessed
by dust, the sub-mm sources represent already a large
fraction of the CIB (between 10 and 60\%).
But the question arises as to determine the relative contribution
of AGN and starburst in dust heating. The fact that the sources
responsible for the X-ray background are not the same as those
for the CIB (Severgnini et al. 2000) does not exclude the possibility
that  the latter are mainly heated by obscured AGN (Barger et al 2001).

ALMA will not suffer confusion. because of its high spatial
resolution (better than 0.1'', see Table \ref{pano}). Its high sensitivity will allow to 
detect more normal (non-ULIRG) objects, such as the  
Lyman-Break Galaxies (Steidel et al 1996, Adelsberger \& Steidel 2000).
Their frequency is of the order of 150/arcmin$^2$   for  z=2.5-3.5,
and this will allow the detection of about 
100 times more submm sources than today. In the millimeter domain,
it is also likely that different objects will be seen than in the optical,
due to extinction (see Melchior, this conference).


\begin{table}[h]
\caption[ ]{Panorama of large ($>$200m$^2$) mm and sub-mm instruments}
\begin{center}
\begin{tabular}{lrcc}  \hline
Name    & Area  &  $\lambda_{min}$  & $\theta$ (``)    \\
\hline
IRAM-30m                & 707  &	1mm	&	10  \\
IRAM-PdB (6x15m)      &1060 &	1mm	&	0.5 \\
NRO-45m          & 1590   &	1.3mm	&	8 \\
NRO       (6x10m)       & 509   &	1.3mm	&	0.5 \\
OVRO  (6x10m)        & 509    &   1.3mm         &               0.5 \\
BIMA   (10x6m)      & 282    &    1.3mm      &                0.5  \\		
CARMA                & 791       &     1.3mm   &	       0.5 \\
SMA   (7x6m)      &200       &0.3mm     &	0.1 \\
GBT-100m        & 7854   & 2.6mm   & 	7 \\
LMT-50m      & 1963 	  & 1mm	 &	6 \\
ALMA (64x12m)  & 7238 &  3-0.3mm  & 0.1-0.01  \\
EVLA (35x25m)  & 17200 &  6mm  & 0.004  \\
\hline
\end{tabular}
\end{center}
\label{pano}
\end{table}

The present submm source counts have been 
compared to theoretical expectations, in particular to 
semi-analytical models, based on the hierarchical scenario for
galaxy formation. Observations help to constrain the numerous
free parameters and to refine the scenarios.
Two strategies are possible: \\
-- either all parameters are fixed according to "realistic" laws or assumptions
(Granato \& Silva 2001) \\
-- or the best fit is searched for, with a restricted number of parameters (Blain 2000).

In particular the models deduce the evolution as a function of 
redshift of the energy released by the baryons through mergers.
The result is that mergers are much more efficient 
for star formation in the past, which could be explained by
the larger gas fraction, and the smaller dynamical time.

\begin{figure}
\centerline{
\psfig{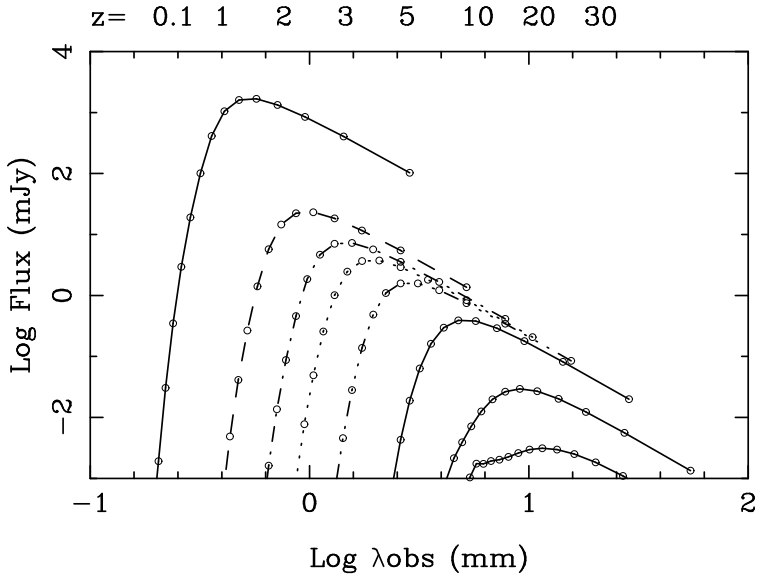}
}
\caption{ Predicted flux for an homogeneous cloud model at T$_k$ = 50K, 
n$_{H2}$ = 10$^3$ cm$^{-3}$ and N$_{CO}$ = 3 10$^{20}$ cm$^{-2}$, for various
redshifts $z$ = 0.1, 1, 2, 3, 5, 10, 20, 30, and $q_0$ = 0.5.
The flux is predicted for the first CO lines of the rotational ladder, 
materialised each by a circle (they are joined by a line only to guide the eye).
The same power has been assumed for the starburst at all redshift,
and $T_{\rm dust}^6 - T_{\rm bg}^6$ is conserved
(cf Combes et al. 1999).}
\label{cohz5}
\end{figure}

To better determine the molecular gas mass fraction, and
therefore the efficiency of star formation, ALMA is necessary to 
detect the CO lines in high-z galaxies.
The detection of CO lines is also favored by the fact that the
high J lines have larger fluxes, and once redshifted, they enter the
submm and mm domain, but the K-correction is not negative, and 
it is more difficult to detect the lines than the continuum at high redshift.
 The prediction of the CO lines flux depends much on the line excitation,
and therefore on the volumic density of clouds n$_{H2}$, the column density
N$_{CO}$ and the temperatureT$_k$. Several simple
models with one or two components have been computed 
(e.g. Combes et al. 1999). One example is shown in figure \ref{cohz5}.

\begin{figure}
\centerline{
\psfig{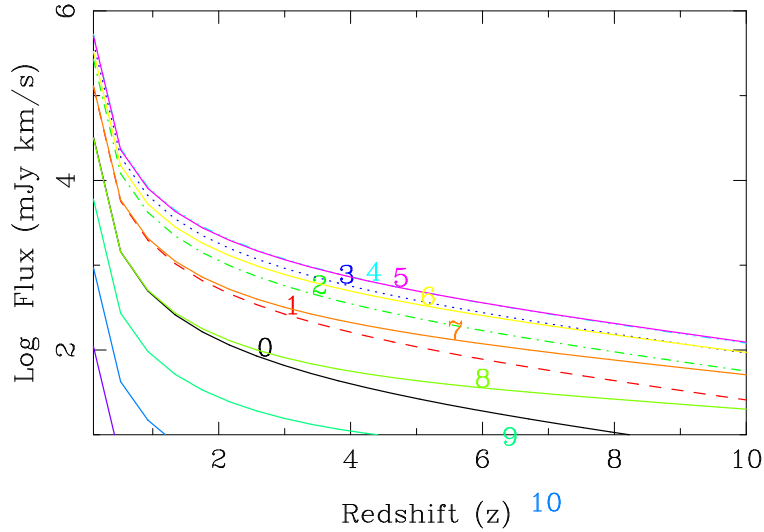}
}
\caption{ Predicted flux as a function of redshift
for the same model as in figure 2, displayed now
for each of the CO lines separately; this shows that the high-J lines are not
much helpful after J=4.}
\label{sdez5}
\end{figure}

Due to the abundance of starbursts at high z, it is likely that most galaxies
will reveal dense and hot molecular gas, and be easily detected from
high J lines. However, it is not necessary to go at very high frequency
since the maximum flux is already reached at J=4-5 (see figure \ref{sdez5}
and e.g. Papadopoulos \& Ivison 2001).

\section{Molecular component, chemistry and dynamics in galaxies}

Other important issues where ALMA will make a breakthrough are:
\begin{itemize}
\item molecular line absorptions: many more continuum sources will be detected
in the mm and submm with the enhanced sensitivity of ALMA; probing 
high column density along the line of sight of remote quasars will allow tackling
the fate of baryons as a function of z, high redshift chemistry, etc... 
(Combes \& Wiklind 1999)

\item the detection of individual molecular clouds in many nearby galaxies 
will allow, through application of the virial theorem, a more accurate determination of
the ill-known CO/H$_2$ conversion factor 

\item the detection of CO at large distance from the center in nearby galaxies,
and the exploration of the outer parts of galaxies

\item dynamics of galaxy centers with high resolution, CO as only tracer where HI 
is deficient (nuclear bars and spirals), and relation with AGN

\item detection of the molecular torus, expected in AGN unification theory

\item dwarf galaxies, and molecular gas outside galaxies (tidal dwarfs, 
star formation complexes...)
\end{itemize}

\section{Conclusions}

ALMA will bring considerable progress in the understanding of
the molecular component and star formation efficiency in nearby
galaxies. One of the main driver is to open the window on the formation
of galaxies: the star formation history, the efficiency as a function of z
to make stars and energy from baryons in mergers, the history
of gas enrichment, the scenario for bulge and disk formation, etc..
The submm and mm domains are a necessary complement of other
wavelength studies, in particular the visible domain, 
since starbursting galaxies suffer significant obscuration.

\vspace*{-0.3cm}


\end{document}